\documentclass[showpacs,amsmath,amssymb,twocolumn,aps]{revtex4}

\usepackage{graphicx}% Include figure files
\usepackage{epsfig}
\usepackage{pstricks}
\usepackage{psfrag}
\usepackage{dcolumn}% Align table columns on decimal point
\usepackage{bm}% bold math

 \def\comment#1{}
 \def\mn#1{*{\marginpar{*\footnotesize #1}}}
 \def\mn#1{}

\begin{document}

\title{
Triangular Lattice Model of Two-Dimensional Defect Melting
}

\author{J\"urgen Dietel}
\email{dietel@physik.fu-berlin.de}
\affiliation{Institut f\"ur Theoretische Physik,
Freie Universit\"at Berlin, Arnimallee 14, D-14195 Berlin, Germany}
\author{Hagen Kleinert}
\email{kleinert@physik.fu-berlin.de}
\affiliation{Institut f\"ur Theoretische Physik,
Freie Universit\"at Berlin, Arnimallee 14, D-14195 Berlin, Germany}
\date{Received \today}

\begin{abstract}
We set up a harmonic lattice model
for 2D defect melting which, in contrast to
earlier simple-cubic models, lives on a triangular lattice.
Integer-valued plastic
defect gauge fields allow for the
thermal generation of
 dislocations and disclinations.
%To this end we first discretize the continuous derivates
%with respect to the x- and y-direction showing that for
% lattices with a cyclic edge vector symmetry requirements are sufficient
%to get unique lattice derivates.
%With the help of these we derive the lattice Hamiltonian of the
%triangular lattice starting from continuum elasticity theory.
%From this theory we derive
The model produces
universal formulas for the  melting temperature
expressed in terms of the elastic constants, which are different
from those  derived for square
lattices. They determine
a Lindemann-like parameter for two-dimensional
melting.
In contrast to
the square crystal which underwent a first-order melting transition, the
 triangular  model melts in two steps.
Our results
are applied to the melting of
 Lennard-Jones and
electron lattices.
\end{abstract}


\maketitle

\section{Introduction}
Melting transitions \cite{shoc1,GFCM2,kleinert118} are
of both technical and theoretical interest.
Under suitable experimental
circumstances
one can study
the melting process in
two dimensions
(2D). If the interparticle interactions are simple, the
lowest-energy crystalline order in 2D is
a close-packed
 triangular. It is
observable
for
electrons on a liquid-helium surface \cite{grimes1},
for adsorbed noble-gas
atoms floating on an incommensurate substrate \cite{heiney1},
for colloidal suspensions of highly charged spheres
confined between parallel glass plates
\cite{murray1},
for paramagnetic colloidal crystals \cite{zahn1},
and for vortex lattices in
thin films \cite{gammel1}.
Simulation experiments
 provide us with further examples:
rigid disks \cite{alder1}, Gaussian cores \cite{stillinger1}, and
particles interacting with  Lennard-Jones \cite{abraham1} as well as
Coulomb forces (Wigner lattice) \cite{bonsall1}.

There are various theories of 2D melting
\cite{halperin1,chui1,ramakrishnan1,SINGLE,TWO,GFCM2,kleinert118,kleinert86}.
% \cite{halperin1}.
% \cite{chui1}.
% \cite{ramakrishnan1}.
% \cite{SINGLE}.
% \cite{TWO}.
% \cite{GFCM2}.
% \cite{kleinert118}
% \cite{kleinert86}.
Most popular is
 the phenomenological theory of Halperin, Nelson, and Young \cite{halperin1}
which, inspired by
the Kosterlitz and Thouless (KT) theory of
vortices in superfluid films \cite{kosterlitz1},
explains the transition by the statistical behavior of defects.
 This KTHNY theory
 suggests
that  melting could proceed in a sequence of two
continuous KT transitions, the first caused by the unbinding
of dislocations, the second by the unbinding of disclinations.
Such a sequence
was indeed
 found in some 2D crystals experimentally and by simulations.

There are also
explicit models on  square lattices which allow to study
defect melting in detail.
These models are minimal in the sense that they
contain the correct harmonic elastic energy
and no anharmonic terms. The defects arise
from
integer-valued plastic gauge fields
which account for
the thermal creation and annihilation of
dislocations and disclinations.
The simplest such model
shows only a single
 first-order melting transition
\cite{SINGLE}. Only after
introducing an extra higher-gradient
elastic term which gives the crystal
a tunable angular stiffness,
the first-order transition
separates into two successive
KT transitions \cite{TWO}.
This splitting can be
observed
in the laboratory, for instance in
Xenon overlayer lattices on graphite \cite{thomy1}.

Presently investigated systems live mostly on  triangular lattices, and these
seem to
melt in  two steps \cite{murray1,zahn1,alder1,muto1,he1,udink1,chen1,gammel1},
 except for the Gaussian core model.
The question arises whether
a minimal melting model
living on a triangular lattice will always
undergo two successive transitions
at the level of first gradient  elastic energies only, so that no
extra higher-gradient terms are needed for splitting them as on square lattices.
It is the purpose of this paper to
answer this question to the affirmative.
We shall generalize
the previously constructed
lattice models
from square lattices
\cite{SINGLE,TWO,GFCM2,kleinert118}
to
triangular lattices and find, indeed, strong indications
that they melt
in two KT transitions.

A universal  feature of all such defect models is
that
it is the combined separation of
dislocations and disclinations
which makes the melting
transition first-order \cite{SINGLE}.
Dislocations alone would cause
a similar phase transition as
in superfluid helium, which
undergoes a continuous transition.

An important virtue of the lattice defect models
on square lattices was that they lead to
a simple universal melting formula
 \cite{GFCM2} determining
the melting point in terms of the elastic constants.
The result is obtained from a lowest-order approximation,
which combines the high-temperature
expansion of the defect contributions
to the free energy density
with the low-temperature expansion.
The  melting temperature
is determined
by
the
intersection of the two curves.
This turns out to
produce precisely the
time-honored Lindemann criterion.
In addition, however,
there is a {\em prediction\/}
 of the universal value of the Lindemann number
where melting occurs.
The accuracy of the approximation
was demonstrated quantitatively for square lattices in 3D and 2D
\cite{GFCM2, kleinert2D}. Recently, the results were
successfully extended to
face-centered and body-centered cubic lattices in 3D
\cite{kleinert03}.
           We shall find a similar
formula for
triangular lattices in 2D.
The  minimal lattice defect model will
be constructed
in Section~II. From its
 partition function
we calculate in  Section~III the melting temperature.
In Section~IV, we compare our result with the
melting transition of  Lennard-Jones and electron lattices.
Section~V discusses the
melting values  when taking into account nonlinear elasticity effects.
In Section~VI we use the model to {\em predict\/}
a universal generalized Lindemann number \cite{lindemann1} for 2D
lattices.

\section{Lattice Hamiltonian with Defects}

The elastic energy of a crystal
in the continuum approximation is given in $D$ dimensions by \cite{landau1}
\begin{equation}
E_{\rm el}=\int d^D x \,\left[
\frac{\mu}{4}(\partial_i u_j+\partial_j u_i)^2
 +\frac{\lambda}{2}
 (\partial_i u_i)^2
     \right] ,   \label{10}
\end{equation}
where  repeated indices are
summed.
The elastic constant
$ \mu $ is the shear modulus, and
$ \lambda $  the Lam\'e constant.
In two dimensions, these constants are related
to the elastic constants
$c_{ij}$  with
lattice symmetry by
$ \mu=c_{66} $ and
$ \lambda= c_{11}-2 c_{66} $.
The combination $c_{11}- c_{66}  $ is the
modulus of
compression.
The fields
 $u_i(x)~(i=x,y)$
are the cartesian components of the
atomic displacements.

In a first step we
 construct a Hamiltonian on a 2D triangular lattice whose
continuum elastic energy agrees with (\ref{10}).
Of course,
this construction is not
unique. %%weiter
We restrict the freedom
by
the following  extension requirements:
\begin{enumerate}
\item
 The lattice derivatives
$ \nabla_x $ and $ \nabla_y $ must have the continuum limits
$\partial_x $ and $\partial_y $, respectively.
\item
  $\nabla_x$, $\nabla_y$
should be  maximally symmetric linear combinations of the
nearest neighbor lattice derivatives
$\nabla_{(l)}$, $l=1,2,3$
with respect to the
point group of the lattice.
\item
$ \nabla_x $ and $ \nabla_y $
should have the same transformation properties
 as $\partial _x $ and $\partial _y $ under the
action of the point group of the lattice.
\end{enumerate}

In the following we denote
 by ${\bf e}_{(l)}$ the  {\it oriented link vectors} of the triangular
lattice which surround a triangular face of the lattice.
There are three such
vectors:
\begin{eqnarray}
{\bf e}_{(1)}  & =&
\left(\cos\frac{2\pi}{6},\sin\frac{2\pi}{6}\right)
 ,\qquad  {\bf e}_{(2)}= (-1,0),       \nonumber \\
{\bf e}_{(3)} & = &\left(\cos\frac{2\pi}{6},-\sin\frac{2\pi}{6}\right)
.   \label{20}
\end{eqnarray}
and three conjugate vectors $ \overline{{\bf e}}_{(l)}=
-{\bf e}_{(l)} $

For each of the  link vectors
there exist a
lattice derivative defined by
\begin{eqnarray}
\nabla _{(1)}f({\bf x}) & = &
\left[f({\bf x}+{\bf e}_{(1)})-f({\bf x})\right]/a  \,, \nonumber \\
\nabla _{(2)}f({\bf x}) & = &
\left[f({\bf x})-f({\bf x}-{\bf e}_{(2)})\right]/a  \,, \nonumber \\
\nabla _{(3)}f({\bf x}) & = &
\left[f({\bf x}-{\bf e}_{(2)})-f({\bf x}+{\bf e}_{(1)})\right]/a \,.
\label{25}
\end{eqnarray}
where $a$ is the lattice spacing.
In addition, there is
 a
conjugate derivative
$ \overline{\nabla}_{(l)}f({\bf x}) $  defined by
$
 -\nabla_{(l)}f({\bf x}) $ as in Eq.~(\ref{25}), but with
the replacement
$ {\bf e}_{(i)} \rightarrow  \overline{{\bf e}}_{(i)}$. From the definition
follows the
 identity
\begin{equation}
 (\nabla_{(1)} +\nabla_{(2)}+\nabla_{(3)}) f({\bf x}) =0  \label{30}
\end{equation}
for any function $ f({\bf x}) $ on the lattice sites.

The 2D point group around each lattice face is given
by $ C_{3v} $ \cite{inui1}. This group causes
permutations of the link directions
$ {\bf e}_{(l)}$ and  the derivates
$\nabla_{(l)}$. These can be represented by matrices
which form a $3\times 3$-dimensional
representation of the point group $ C_{3v} $ in the space of
linear combinations of the link vectors or of the lattice derivates.

We now define cartesian lattice versions
 $ \nabla_x $ and $ \nabla_y $
of the
derivatives $ \partial _x$ and $\partial _y$
as
$ C_{3v} $-
symmetric
linear
combinations of the lattice derivatives
$\nabla_{(l)}$:
\begin{eqnarray}
\!\!\!\!\!\! \partial _{x,y}  \!\!\! & \rightarrow &
\!\!\!\nabla_{x,y}\equiv \frac{2}{3} \,
{{e}_{(l)x,y}}{}_{}
 \nabla_{(l)}\, .
           \label{40}
\end{eqnarray}
Inserting the components from (\ref{20})
these become
\begin{eqnarray}
\!\!\!\!\!\!\partial _{x}  \! & \! \rightarrow & \!
 \nabla_{x}\equiv  \frac{2}{3} \!
\left[\!\left(\frac{1}{2}\nabla_{(1)}\!-\!\nabla_{(2)}
\!+\!\frac{1}{2}\nabla_{(3)}\right)
 \right] \! ,~ ~~     \nonumber           \\
\partial _{y}\!& \! \rightarrow &
  \nabla_{y}\equiv \frac{2}{3}\!
\left[\!\left(\! \frac{\sqrt{3}}{2}\nabla_{(1)}
\!-\!  \frac{\sqrt{3}}{2} \nabla_{(3)}\right)
\right]\!.        ~~~    \label{50}
\end{eqnarray}
It is easy to check that these
fulfill the above requirements~$1$--$3$.

Without proof we mention that by using group theoretical  methods
(see e.g. \cite{inui1}) we can show
 that
the extension requirements $ 1-3 $ are sufficient to obtain unique lattice
derivates $ \nabla_x $ and $ \nabla_y $. This is true for all other
lattice symmetries. The only property needed for the proof is
that the lattice possesses a  link vector which generates all others
by applying the symmetry group of the lattice.

\section{Lattice Hamiltonian}
It is now easy to set up the lattice version
of the continuum Hamiltonian
(\ref{10}).
We simply replace the partial derivatives
by the lattice derivatives (\ref{50}),
and obtain
\begin{align}
& E_{\rm lat}  =  \frac{v}{2}
\sum_{{\bf x}}
 \left\{
\frac{\mu}{2}[\nabla_i u_j(x)+\nabla_j u_i(x)]^2
 +\lambda [\nabla_i u_i(x)]^2
 \right\}.
       \label{80}
\end{align}
where the sum $ \sum_{{\bf x}} $ runs over all lattice sites, and
$ v=\sqrt{3}a^2/2 $
is the
area of the
fundamental cell of the lattice
(for the square lattice
$ v=a^2 $).
As in the earlier work
on square lattices
\cite{GFCM2}
 we shall first consider a symmetrized
version of the Hamiltonian (\ref{80}) which has the advantage
of leading to
 explicit formulas for the melting transition for all
$ \mu$, $ \lambda$.
This arises by replacing $\nabla_{i} $
by $\overline\nabla_i$ in the $ \lambda $ term:
\begin{align}
& E_{\rm lat}  \approx   \frac{v}{2}
\sum_{{\bf x}}
\left\{
\frac{\mu}{2}[\nabla_i u_j(x)+\nabla_j u_i(x)]^2
 +\lambda [\overline{\nabla}_i u_i(x)]^2
 \right\}.
       \label{85}
\end{align}
Without an explicite
discussion in this paper we mention that one can show the difference in
the results using the symmetrized Hamiltonian (\ref{85})
and the Hamiltonian (\ref{80}) to be negligible
for triangular as well as square lattices.
The most symmetric way of writing (\ref{80})
is
\begin{align}
& E_{\rm lat}\!\! = \!\! \frac{\sqrt{3}a^2}{4}
\sum_{{\bf x}}\Bigg\{
\frac{4}{9} \, c_{11} \Bigg(\!\sum_j \nabla_{(j)} u_{(j)}
\!-\frac{1}{2}\sum_{i \not= j} \nabla_{(i)} u_{(j)}\! \Bigg)^2
\nonumber \\
&
+ \frac{1}{3} \, c_{66}\Bigg( \sum_{k ij} \epsilon_{kij}
\nabla_{(i)}  u_{(j)} \Bigg)^2     \nonumber \\
&
+\frac{2}{3} \, c_{66}  \bigg(
\sum_{k \not= i \not= j} 2 \nabla_{(k)}  u_{(k)} \nabla_{(i)}  u_{(j)}- 2
\nabla_{(i)}  u_{(k)} \nabla_{(k)}  u_{(j)}  \nonumber \\
& -\nabla_{(k)}  u_{(k)} \nabla_{(i)}  u_{(i)}+
\nabla_{(i)}  u_{(k)} \nabla_{(k)}  u_{(i)}\bigg)
\Bigg\} .        \label{90}
\end{align}
 Here we have expanded the displacement vector
in (\ref{80})
in a symmetric way to
${\bf u}(x)= u_{(l)}({\bf x}) {\bf e}_{(l)}$.
This representation of $ {\bf u}({\bf x}) $ is not unique
due to the overcompleteness of the three link vectors
(\ref{20}). It becomes unique
by setting
$ u_{(3)}=0 $ and taking into account (\ref{30}), yielding
\begin{align}
& E_{\rm lat}\!\! = \!\! \frac{\sqrt{3}a^2}{4}
\sum_{{\bf x}}\Bigg\{
 c_{11} \bigg(
\nabla_{(1)} u_{(1)} +\nabla_{(2)} u_{(2)} \bigg)^2
\nonumber \\
&
+ \frac{1}{3} \, c_{66}\Bigg(
 \nabla_{(1)} u_{(1)} -\nabla_{(2)} u_{(2)}  +2
\nabla_{(2)} u_{(1)}- 2 \nabla_{(1)} u_{(2)} \Bigg)^2     \nonumber \\
&
+4 \, c_{66}  \bigg( \nabla_{(1)} u_{(2)} \nabla_{(2)} u_{(1)}-
  \nabla_{(1)} u_{(1)} \nabla_{(2)} u_{(2)}  \bigg)  \Bigg\}  \,.
      \label{95}
\end{align}

\section{Including Defect Gauge Fields}
With the goal of studying
defect-induced melting transitions,
the lattice representations (\ref{80}) and (\ref{85})
of linear elasticity must be extended
by integer-valued defect gauge fields, the lattice version of
plastic fields.
From the textbook \cite{GFCM2}
we know how to do this for square lattices.
The defect gauge fields
reflect the fact
that
due to fluctuations,
atoms are capable of exchanging positions with their neighbors
and migrate eventually
through the entire crystal. This
process of self-diffusion makes it impossible to specify the displacement
field uniquely. Thus, as a matter of principle, the displacement field is
multi-valued. It is
determined only up to an arbitrary lattice vector,
since it is impossible to say
whether an atom is displaced by ${\bf u}({\bf x})$ or by
$ {\bf u}({\bf x})+ a N_{(l)}({\bf x}) {\bf e}_{(l)} $
where $N_{(l)}({\bf x})$ is integer.
The displacement vectors may have jumps across so-called
{\em Volterra surfaces\/}
which are described by integer-valued defect
gauge fields
$n_{(lm)}({\bf x})$  in the
 lattice model \cite{GFCM2}.  These are, in general,
not symmetric.

The defect gauge fields are included into the lattice derivatives of
the Hamiltonian (\ref{80}) by
the replacement
\begin{eqnarray}
 \nabla_{x,y} {\bf u}({\bf x}) &   \rightarrow &
\frac{2}{3} {{e}_{(l)}}_{x,y}  \label{120}
%\\&   \times&
\left[\nabla_{(l)}  {\bf u}({\bf x})- \,n_{(lm)}({\bf x})
{\bf e}_{(m)}  \right] \!.~~~
\end{eqnarray}
The classical partition functions of the triangular crystal
including dislocation and disclination degrees of freedom
is then given by
\begin{equation}
Z= \prod_{{\bf x},i} \left[  \int_{-\infty}^\infty
  \frac{d u_i({\bf x})}{a} \right]
\sum_{\{n_{{lm}}({\bf x})\}} \Phi[n_{(lm)}] \
\exp[-E_{\rm lat}/k_B T].
\label{130}
\end{equation}
For simplicity, we choose periodic boundary conditions for the displacement fields
$ {\bf u}({\bf x}) $.
Since defects are spanned
over an integer number of fundamental cells of the lattice,
the functional $\Phi[n_{(lm)}]$ should restrict the range
of the integer-valued gauge
field $ n_{(lm)} $. Furthermore,
it acts also as a gauge-fixing, in such a way that only physically independent
degrees of freedom are summed.
% One
% restriction reflects the fact
% each  defect
% is associated with
%  a fundamental cell of the lattice. \mn{is this what you want ot say?}
%  Another restriction
% is necessary to
%  fix the gauge degrees of freedom in the gauge field
% $n_{(lm)}$,
% so that only physically independent
% degrees of freedom appear in the partition function.

Due to the identity (\ref{30}) we have
the freedom to carry out the replacement
$ \nabla_{x,y} \rightarrow
\nabla_{x,y}+a_{x,y} (\nabla_{(1)} +\nabla_{(2)}+\nabla_{(3)})  $
in (\ref{50}) with some constants $ a_{x,y} $
without changing the elastic energy in (\ref{80}) or
(\ref{85}). One can see that by considering
relation (\ref{120}), the resulting Hamiltonian and also
the partition functions depends strongly on $ a_{x,y} $.
This is because the replacement (\ref{120})
does not take into account the
fact that defect lines as dislocations and disclinations
are built of interstitials or vacancies, respectively, which allows only
one
interstitial or vacancy per fundamental cell. This
restricts the
{\em Volterra surfaces} to run
through  an integer number of fundamental cells.
 \mn{explain} (See Fig.~1).
This restriction on the defect gauge fields
results in the following constraint
\begin{equation}
 n_{(1l)}({\bf x})+ n_{(2l)}({\bf x})+n_{(3l)}({\bf x})=0 ,  \label{140}
\end{equation}
for all $ l=1,2,3 $.
Here we take into account
that
 for a dislocation
the
jump  over the
Volterra surface is constant.
Disclinations are built of dislocations within our lattice model.
By taking into account this constraint
on the defect gauge fields the resulting partition function
does not longer depend on the replacement constants $ a_{x,y} $. \\

We now must identify and eliminate the gauge degrees of freedom
in Eq. (\ref{130}).
The overcompleteness of the three basis vectors
$ {\bf e}_{(l)} $ of the two-dimensional lattice implies
that the replacement (\ref{120}) leads
to
the same
defects if the integer numbers
satisfy
$ n_{(li)}({\bf x}){\bf e}_{(i)}=0 $ for $ l=1,2,3$.
The rest of the gauge freedoms can be eliminated by analogy
with the gauge fixing on the square lattice in
the textbook~\cite{GFCM2}.
Initially, the displacement fields cover only a single fundamental cell.
We can extend the range to the entire crystal by fixing
 two of the last four components of the defect gauge field $  n_{(lm)} $.\mn{of what??}
One more component  \mn{which??} can be eliminated since
$  E_{\rm lat} $ depends
on the derivates $ \nabla_x u_y $
and $ \nabla_y u_x $ via the sum $  \nabla_y u_x+ \nabla_x u_y $.
\begin{figure}
% \vspace*{-1.5cm}
\hspace*{0.5cm}
\begin{center}
\includegraphics{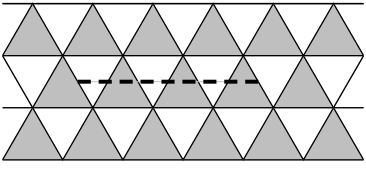}
% \vspace*{-2.4cm}
% \hspace*{-0.5cm}
\caption{Cut in a triangular lattice.
For each lattice site, the shaded regions show the faces which are surrounded
by the oriented link vectors (\ref{20})
associated with
the lattice
derivates $ \nabla_{(l)} $ in Eq.~(\ref{25}). Dashed line indicates
 a {\it Volterra surface} crossing an integer number of
fundamental cells.}
\end{center}
\vspace*{-0.5cm}
\hspace*{0.5cm}
\end{figure}

Finally, we must take  care
of the
 constraints due to the periodic boundary conditions.
The technical aspects of the elimination of
the gauge degrees of freedom are deferred to Appendix A.
Eventually we arrive
at the partition function:
\begin{align}
 & Z  =   \prod_{{\bf x},i}\left[
 \int_{-\infty}^\infty\frac{u_i({\bf x})}{a} \sum_{n({\bf x})=-\infty}^{\infty}
\right]
                  \label{170} \\
 &  \!  \times \! \left((1-\delta_{{\bf x},{\cal B}})+\delta_{{\bf x},{\cal B}}
 \!  \!\!\! \!\! \! \! \!\sum_{n_{(11)}({\bf x}),n_{(21)}({\bf x})=-\infty}^{\infty}
 \right)
 \exp[-E_{\rm lat}/k_B T]
\,,    \nonumber
\end{align}
with the lattice energy
(\ref{80})  and the replacements of the lattice derivatives
\begin{align}  \!\!\!\!\!\!\!
&  (\nabla_{x},\nabla_{y}) \times  {\bf u}({\bf x})
 \rightarrow  \frac{2}{3 }
 ({{e}_{(l)}}_{x}, {{e}_{(l)}}_{y}) \times
 \nabla_{(l)}  {\bf u}({\bf x})          \label{180}      \\
&~~~~~~~~~~~~~~~~~  + \left( \begin{array}{cc}
0  & \frac{1}{ \sqrt{3}} \,  n({\bf x})
   \\
\frac{1}{ \sqrt{3}} \, n({\bf x})&  0
\end{array} \right)      \nonumber  \\
& ~~~~~+
\delta_{{\bf x},{\cal B}}  \,
 \frac{1}{2} \left(
\begin{array}{cc}
n_{(21)} ({\bf x})  &
-\frac{1}{\sqrt{3}} \, n_{(21)} ({\bf x}) -\frac{2}{\sqrt{3}}  n_{(11)}
   \\
+ \sqrt{3}
n_{(21)} ({\bf x}) &   -2 n_{(11)} ({\bf x})-n_{(21)} ({\bf x})
\end{array} \right)   \,.         \nonumber
\end{align}
where $ {\cal B} $ contains the sites at the rightmost
and top boundary of the sample.
%Of course, we could choose other boundaries
%since the periodic boundary conditions of the displacement fields
%$ {\bf u}({\bf x}) $ imply that
%the model has no physical boundary at all.

The gauge degrees of freedoms are fixed by choosing
$ n({\bf x})=0 $ for ${\bf x} \in {\cal B} $, further
  $ \nabla_x  n({\bf x})=0 $
on the rightmost  boundary $ {\cal B}_r $,
 and  $ \nabla_y  n({\bf x})=0 $ on the
topmost boundary $ {\cal B}_t $ of the model. This will be summarized by
the statement
\begin{equation}
 n({\bf x})=0   \quad \mbox{for} \quad {\bf x} \in {\cal B}'\,. \label{185}
\end{equation}
where $  {\cal B}' $ is the extended boundary consisting
on the sites $ {\cal B} $,
$ {\cal B}_t+a {\bf e}_{(1)}  $ and $ {\cal B}_r+a {\bf e}_{(2)}  $.
The boundary fields $ n_{(11)}({\bf x}) $and $ n_{(21)}({\bf x}) $
 in (\ref{180})
result from the periodic boundary conditions of the
displacement field $ {\bf u} $,
which restrict the gauge degrees of freedom.

The final partition function
(\ref{170})
describes
a triangular crystal
 with harmonic
elastic  fluctuations and
fluctuating dislocations and disclinations.

By a standard duality transformation
it is possible to rewrite
this partition function in a canonical form.
Introducing the conjugate
 stress fields $ \sigma _{ij}({\bf x})$
via
an auxiliary integral, we see that (\ref{170})
is  equal to
\begin{align}
& Z =  \left[\frac{\mu}{4(\lambda+\mu)}\right]^{N/2}
\left(\frac{1}{2\pi\beta_{\triangle}}\right)^{3N/2}
\prod_{{\bf x}, i\leq j}\!
\left[ \int_{-\infty}^\infty d\sigma_{ij}\right] \nonumber \\
& \times \prod_{{\bf x}}\left[ \sum_{n({\bf x})=-\infty}^{\infty}
\int_{-\infty}^\infty\frac{d {\bf u}({\bf x})}{a}
\right]
\exp\Bigg\{-\frac{1}{2 \beta}
\sum_{{\bf x}} \nonumber \\
&   ~~~~~~~~ \times
 \Bigg[\sum_{i<j}\sigma _{ij}^2+
\frac{1}{2}\sum_i \sigma _{ii}^2-
\frac{\lambda}{4(\lambda+\mu)}\Big(\sum_i \sigma _{ii}
\Big)^2\Bigg]\Bigg\}  \nonumber \\
&\times \exp\Bigg[i \, 2 \pi \sum_{{\bf x}}
\frac{2}{3}\bigg[\sum_{i} {e}_{(l)}{}_{i}
 \nabla_{(l)}  u_i \,\sigma_{ii}         \label{190}  \\
&  + \left({{e}_{(l)}}_{1}
 \nabla_{(l)}  u_2+ {{e}_{(l)}}_{2}
 \nabla_{(l)}  u_1\right)\sigma_{12} \bigg]
 + H_{\sigma n} \Bigg]  \nonumber
\end{align}
where  $\beta_{\triangle}\equiv a^2 \mu \sqrt{3}/2 k_B T(2\pi)^2$.
For details see \cite{GFCM2}.

The energy $ H_{\sigma n} $ couples the stress field $ \sigma _{ij} $
to the integer value fields $ n_{(ij)}({\bf x}) $:
\begin{align}
& H_{\sigma n} =i \, 2 \pi \sum_{{\bf x}}
\frac{2}{\sqrt{3}} \, \sigma_{12} \, n({\bf x}) +
 \ln \Bigg[ \prod_{{\bf x} \in {\cal B}}\Bigg[ \sum_{n_{(11)}, n_{(21)}({\bf x})=-\infty}^{\infty}
\Bigg] \nonumber \\
&
  \exp\bigg[i \, 2\pi
\bigg(\frac{1}{2}\,\sigma_{(11)}({\bf x}) n_{(21)}({\bf x})
- \frac{1}{2} \sigma_{(22)}({\bf x})\left(n_{(21)}({\bf x}) + \right.
\nonumber \\
&\left. +2
n_{(11)}({\bf x})\right)
+ \frac{1}{\sqrt{3}}\, \sigma_{(12)}({\bf x})\left(
 n_{(21)}({\bf x})-
n_{(11)}({\bf x})
\right)\bigg) \bigg] \Bigg].   \label{192}
\end{align}
This is the
   dual representation of (\ref{170}).
For very large lattices,
the integer-valued fields  $ n_{(11)}({\bf x}) ,n_{(21)}({\bf x}) $
have, of course,
no physical implications
since
they exists only on the boundary of the system.
For the present calculations on a finite lattice
they are, however, needed.
\comment{
We will show in the following
that this is only  true
in the fluid phase.
The reason is that the physical relevant lowest energy
excitations in the solid phase are basic dislocations and disclinations
which are built in general
of an infinite number of integer number defect fields
$ n({\bf x}) $.
For large systems,
the fields  $ n_{(11)}({\bf x}) ,n_{(21)}({\bf x}) $ do not
have
physical implications
since
they exists only on the boundary of the system. When discussing }
 \mn{still do not agree}
%For the  We shall see
%that this is not true for the triangular lattice in the
%solid phase but one can neglect these surface terms in the fluid phase.

The partition function (\ref{190}) corresponds
to the Hamiltonian (\ref{80}) including defects. The partition
function for the symmetrized Hamiltonian (\ref{85}) is given by the
same expression with the replacement
\begin{equation}
\frac{\lambda}{4(\lambda+\mu)}\Big(\sum_i \sigma _{ii}
\Big)^2 \rightarrow
\frac{\lambda}{4(\lambda+\mu)}\left(\sum_i
\frac{\overline{\nabla}_{i}}{\nabla_i} \, \sigma _{ii}
\right)^2                               \label{195}
\end{equation}
in the third term of the first exponent in (\ref{190}).
In the following sections, we shall use
the symmetrized Hamiltonian (\ref{85})
corresponding to the partition function (\ref{190}) with
(\ref{195}) as the basis for calculating physical quantities.

\section{Melting Transition}
We now determine the melting temperature of the
model.
%As announced in the introduction, we shall do this by
%calculating
%the low- as well as the high-temperature expansions of the
%partition function $ Z $, and determining where they intersect.

\subsection{Lowest-Order Approximation}

It has been shown in the textbook \cite{GFCM2}
that the intersection of the free energies of the
two limiting curves of the high- and low- temperature expansion
yields good estimates for the melting temperatures
for simple cubic lattices in three dimensions
and for square lattices in two dimensions
(see Figs. 12.1 and 12.2 on pp. 1084--1085).

For small $T$, the defects
are frozen out, and
the partition function has the classical limit of the partition function
of the Hamiltonian (\ref{85}) with
the derivate substitution (\ref{120}) and
$ n({\bf x})\equiv 0 $:
 \begin{equation}
Z_{T \to 0}
  =  (2 \pi\beta_{\triangle})^{-N} \left(\frac{\mu}{\lambda+2 \mu
}\right)^{N/2}
e^{ -N \ell_{\triangle}}.           \label{200}
\end{equation}
The parameter $ \ell_{\triangle} $ denotes
 the trace of the logarithm of the triangular
lattice Laplacian divided by the number of sites.
The
Fourier transforms of the lattice derivates (\ref{25}) are
\begin{align}
& K_{(1)}  =  \frac{1}{ia}(e^{i a {\bf k} \cdot{\bf e}_{(1)}}-1)
\quad \,,\qquad
K_{(2)} =  \frac{1}{ia}(1-e^{-i a {\bf k}\cdot{\bf e}_{(2)}})\nonumber \\  &
K_{(3)} =  \frac{1}{ia}
(e^{-i a {\bf k}\cdot{\bf e}_{(2)}}-e^{i a {\bf k}\cdot{\bf e}_{(1)}})  \,.
\label{202}
\end{align}
The conjugate lattice derivatives
have the complex-conjugate
Fourier transforms
$ \overline{K}_{(l)}=K^*_{(l)} $.\mn{OK?}
Using these,
we calculate $\ell_{\triangle}$  from the
momentum  integral
\begin{eqnarray}
\ell_{\triangle}&  = &
\frac{1}{2 A_{\rm BZ}} \int_{\rm BZ}  d^2k
\log\left( \left[\frac{4 a^2 }{9} \overline{K}_{(l)}
K_{(m)}
{\bf e}_{(l)}\cdot
{\bf e}_{(m)} \right]^2\right)
        \nonumber \\
&  \approx &  1.22 \,.
\label{205}
\end{eqnarray}
More explicitly, the argument of the logarithm is
\begin{align}
&\!\!\! \frac{4a^2}{9} \overline{K}_{(l)}
K_{(m)}
{\bf e}_{(l)}\cdot
{\bf e}_{(m)}      \label{210}     \\
&\!\!\! =
\bigg[4-\frac{4}{3}\left(\cos a {\bf k} \cdot {\bf e}_{(1)}+
\cos a {\bf k} \cdot {\bf e}_{(2)}
+ \cos  a {\bf k} \cdot {\bf e}_{(3)} \right) \bigg]
\nonumber  \,.
\end{align}
The momentum integral in (\ref{210}) runs over the
2D Brioullin zone
whose area is
$ A_{\rm BZ} = (2 \pi)^2 \sqrt{3}/2 a^2$.
Let us compare the value (\ref{205})
with the corresponding on the square lattice in Ref.~\cite{GFCM2}, where
 the low-temperature partition
function looks like
(\ref{200}), but  with
$ \ell_{\triangle}\approx1.22$  replaced by $ \ell_{\square}\approx 1.14 $,
 and
$\beta_{\triangle}
\equiv a^2 \mu \sqrt{3}/2 k_B T(2\pi)^2 $ by
$ \beta_{\square} \equiv a^2 \mu/k_B T(2\pi)^2$.

Above the melting point, the partition function is
 calculated from  the
dual representation (\ref{190}).
In the  high-temperature limit,
 the defects are prolific.
The sum over $n({\bf x}) $  enforces $\sigma_{12}({\bf x}) $ to have discrete values
such that terms with
$\sigma_{12}({\bf x})\not= 0$ are exponentially small
 for large
temperature. Thus we can  restrict ourselves
to terms with
$ \sigma_{12}({\bf x})=0 $ only.
 After carrying out
the integrals over the lattice displacements $ {\bf u}({\bf x}) $, we obtain
the constraint that the stress field is divergenceless.
Its independent components
are
integrated out  using  the formulas
\begin{align}
& \prod_{{\bf x}}\left[ \int _{-\infty}^\infty d\sigma_{11}\right]
 \delta\left(
\frac{2a}{3} {e_{(l)}}_{i} \overline{\nabla}_{(l)} \sigma_{i1}({\bf{x}})\right)
\label{212} \\
& =\exp\left[-\frac{N}{2 A_{\rm BZ}} \int_{\rm BZ} d^2 k \log\left(
\frac{4 a^2}{9} \overline{K}_{(l)}
K_{(m)} {e_{(l)}}_x  {e_{(m)}}_x \right)\right]
           \nonumber \\
&  =\exp\left[-\frac{N}{2 A_{\rm BZ}} \int_{\rm BZ} d^2 k \,
 \log\left(2 -2 \cos
a {\bf k} \cdot {\bf e}_{(2)}\right)\right]=1    \nonumber
\end{align}
and
\begin{align}
& \prod_{{\bf x}}\left[
\int_{-\infty}^\infty d\sigma_{22} \right]  \delta\left(
\frac{2a}{3} {e_{(l)}}_{i} \overline{\nabla}_{(l)} \sigma_{i2}({\bf{x}})\right)
\label{215}  \\
& =\exp\left[-\frac{N}{2 A_{\rm BZ}} \int_{\rm BZ} d^2 k \log\left(
\frac{4a^2}{9} \overline{K}_{(l}
K_{(m)} {e_{(l)}}_y  {e_{(m)}}_y \right)\right]
           \nonumber \\
&  =\exp\bigg\{-\frac{N}{2 A_{\rm BZ}} \int _{\rm BZ}d^2 k \,\log\bigg[
\frac{1}{3} \left(6 - 4 \cos a {\bf k} \cdot {\bf e}_1  \right.
\nonumber  \\
& \left.~~~~~~~~~~~\!
 - 4 \cos a {\bf k} \cdot {\bf e}_3 + 2 \cos a {\bf k} \cdot {\bf e}_2 \right)
\bigg]
\bigg\} \approx  \left(\frac{1}{1.15}\right)^N  \,.    \nonumber
\end{align}
The partition function
$ Z $ in this limit is
\begin{equation}
Z_{T \to \infty}
 =  \left(2 \pi \beta_{\triangle}\right)^{-3N/2}
\left[\frac{\mu}{4(\lambda+\mu)}\right]^{N/2}
{C}_{\triangle}^{N/2}        \,,      \label{220}
\end{equation}
where $ {C}_{\triangle} $ is the
constant
\begin{equation}
{C}_{\triangle} \approx 0.57 \,.
      \label{230}
\end{equation}

From the intersection of the high-temperature expansion (\ref{220}) with the
the low-temperature expansion
(\ref{200}) we obtain the lowest-order  result for the
melting transition of
a triangular lattice
\begin{equation}
\beta_{\triangle} (1+\nu) \approx
\frac{1}{4 \pi} e^{2 \ell_{\triangle}}{C}_{\triangle} \,
 \approx 0.52
\label{240}
\end{equation}
where $ (1+\nu) =2(\lambda+\mu) /(\lambda+2 \mu)$.
This is somewhat smaller than
 $ \beta_{\square} (1+\nu) \approx 0.81 $ for the square lattice \cite{GFCM2}.
%The reason turns out to be
%the more violent fluctuations on a triangular lattice compared
%to the square lattice.

Let us now see how these results
are changed by a systematic
improvement of the
lowest-order
low- and high-temperature
expansions
(\ref{200}) and (\ref{220}).

\subsection{Beyond the Lowest Order}
In the  low-temperature
regime,
we  include
the leading contributions with
nonzero
 defect gauge fields $n({\bf x})$ in the partition function
(\ref{170}). In the dual representation
(\ref{190}), we must
take into account the
integer nature of $ n({\bf x}) $.
This was done in Ref. \cite{GFCM2} by
reexpressing (\ref{190})
in terms
of the physical
defect density field, which is
the double-curl of the defect gauge field.
Here we carry out a similar calculation for the triangular lattice.
After performing,
in (\ref{190}),
 the integrals
over the displacement field $ {\bf u}({\bf x}) $ and the stress fields
$ \sigma_{ij}({\bf x}) $
we obtain the improved
low-temperature approximation
$ Z=Z_{T \to  0} Z^{\rm def}_{\rm \triangle}  $ with
the defect correction factor
\begin{align}
& Z_{\rm \triangle}^{\rm def} = \prod_{{\bf x}}
\sum_{n({\bf x})=-\infty}^{\infty}
\left((1-\delta_{{\bf x},{\cal B}})+\delta_{{\bf x},{\cal B}}
 \sum_{n_{(12)},n_{(11)}({\bf x})=-\infty}^{\infty} \right)
 \nonumber \\
&
\times
\exp\Bigg[-4\pi^2 \beta_{\triangle} (1+\nu) \sum_{{\bf x}, {\bf x}'}
\frac{4a^2}{9} \, \nabla_{(l)} e_{(l)}{}_x
\nabla_{(m)}e_{(m)}{}_y \, \tilde{n}({\bf x})
\nonumber \\
& ~~~~\times v_{\rm \triangle}({\bf x}-{\bf x}')
\; \frac{4a^2}{9}\nabla_{(k)}e_{(k)}{}_x
\nabla_{(n)}e_{(n)}{}_y
\, \tilde{n}({\bf x}')\Bigg] .  \label{250}
\end{align}
and
\begin{align}
& \tilde{n}({\bf x})  =  \frac{2}{\sqrt{3}} \,n({\bf x})
+\left( \frac{\nabla_{(l)} {e_{(l)}}_x}
 { \nabla_{(l)} {e_{(l)}}_y} - \frac{1}{\sqrt{3}} \right)
n_{(11)}({\bf x}) \, \delta_{{\bf x},{\cal B}}
 \label{252}  \\
&
+
\left(\frac{1}{\sqrt{3}}-
 \frac{1}{2} \frac{\nabla_{(l)} {e_{(l)}}_y}{ \nabla_{(l)} {e_{(l)}}_x}
+\frac{1}{2} \frac{\nabla_{(l)} {e_{(l)}}_x}{ \nabla_{(l)} {e_{(l)}}_y}
\right)
 n_{(21)}({\bf x}) \,
\delta_{{\bf x},{\cal B}}
    \,.
\end{align}
The numbers $ n({\bf x}) $ fulfill the boundary condition (\ref{185}).
In this way, we have expressed
the low-temperature corrections
to the harmonic partition
function
as a
partition function of integer-valued fields
$ n({\bf x}) $ in the bulk and two integer valued fields
$ n_{(11)}({\bf x}) $,  $ n_{(21)}({\bf x}) $ on the boundary of the sample.

The interaction potential $ v_{\rm \triangle} $ is given by the inverse
square of the triangular Laplacian
\begin{equation}
 v_{\rm \triangle}({\bf x})=
\left[
\frac{4a^2}{9}
\nabla_{(l)}\nabla_{(m)}{\bf e}_{(l)}\cdot {\bf e}_{(m)}
\right]^{-2}({\bf x})\equiv
\frac{1}{a^4}\, \square^{-2}_{\triangle} ({\bf x}).
       \label{255}
 \end{equation}
It is not easy to obtain the leading-order contributions
of $  Z_{\rm \triangle}^{\rm def} $. This is mainly due to cancellation
effects in the configuration sum coming from the interaction
of the fields $ n({\bf x}) $ and $ n_{(11)}({\bf x}) $, $ n_{(21)}({\bf x}) $
on the boundary of the sample.
A more efficient
calculation is possible if
we go over from the defect gauge fields
to the
{\em defect density field\/}
%s $ \nabla_{(l)}{\bf e}_{(l)}{}_x
%\nabla_{(m)}{\bf e}_{(m)}{}_y \, \tilde{n}({\bf x}) $ which interact
%via $ v_{\rm \triangle} ({\bf x}-{\bf x}')$.
% although the number
%of $ n({\bf x}) $ fields which contribute to this configuration are
%of infinite number for $ N \to \infty $.  This is a well known
%situation for the square lattice \cite{GFCM2}.
%The reason for this cancellation  is because the operator
%$  \nabla_{(l)}{\bf e}_{(l)}{}_x
%\nabla_{(m)}{\bf e}_{(m)}{}_y  $ acting on $ \tilde{n}({\bf x}) $ has a
%non trivial core. Without this operator in (\ref{250}) the boundary
%fields $ n_{(11)}({\bf x}) $, $ n_{(21)}({\bf x}) $ would contribute
%to the partition function $ Z_{\rm \triangle}^{\rm def} $
%only via negligible boundary terms for an infinite system.
\mn{deleted, was ununderstandable}
given by
\begin{equation}
\eta({\bf x})= \frac{2a^2}{3}
\, \epsilon_{kl} \epsilon_{mn} \, \nabla_{k} \nabla_{m} \,
\, {{e}_{(i)}}_ l \, n_{(ij)}({\bf x}) \, {e_{(j)}}_n  \,.   \label{260}
\end{equation}
The  defect field $ \eta({\bf x}) $
describes  disclination degrees
of freedoms \cite{GFCM2}. The dislocations
arise from dipoles of two nearby disclinations.
The main advantage of $ \eta({\bf x}) $ is that
it is invariant under defect gauge transformations of
the integer-valued fields $  n_{(ij)} $.
\begin{figure}% \vspace*{-1.5cm}
\hspace*{-0.5cm}
\begin{center}
\includegraphics{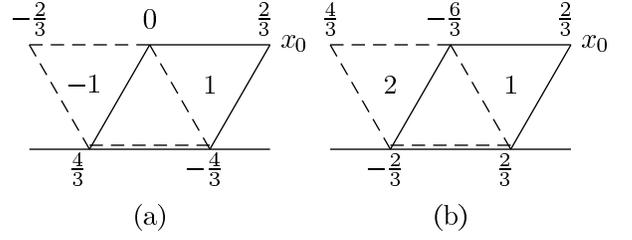}
% \vspace*{-1.3cm}
% \hspace*{-0.5cm}gure}
\caption{Basic configurations
of defect density
$ \eta({\bf x}) $ associated with
the defect gauge field
$ n_{(ij)}({\bf x}) = \delta_{i,1}
\delta_{j,2} \, \delta_{{\bf x},{\bf x}_0} $ (a), and
$ n_{(ij)}({\bf x})=\delta_{i,1} \delta_{j,1} \delta_{{\bf x},
{\bf x}_0} $ (b). The numbers in the triangles
denote the multiplicity of basic charges $ 2/3 $.
 }
\end{center}
\vspace*{-0.5cm}
\hspace*{0.5cm}
\end{figure}
\begin{figure}
% \vspace*{-1.5cm}
% \hspace*{-0.5cm}
\begin{center}
\includegraphics{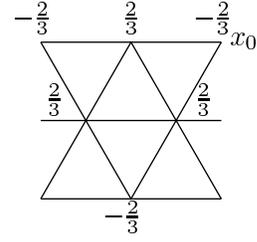}
% \hspace*{-1.3cm}
% \vspace*{-0.5cm}
\caption{Defect configuration $ \eta({\bf x}) $
for
$ n_{(ij)}({\bf x}) = (\delta_{i,2} \delta_{j,1}  - \delta_{i,1}
\delta_{j,2} - \delta_{i,1}
\delta_{j,1})\, \delta_{{\bf x},{\bf x}_0} $.
Its contribution to the free energy is given by the first term in Eq.~(\ref{272}).
 }
\end{center}
\vspace*{-0.5cm}
\hspace*{0.5cm}
\end{figure}
Thus, we obtain for
$ Z_{\triangle}^{\rm def} $ of Eq.~(\ref{250}):
\begin{align}
&Z_{\rm \triangle}^{\rm def}=\prod_{{\bf x}}
\sum_{\eta({\bf x})}
e^{-4\pi^2 \beta_{\triangle} (1+\nu)  \eta({\bf x})
v_{\rm \triangle}({\bf x}-{\bf x}')  \eta({\bf x}´)
\; } .  \label{270}
\end{align}
The sum $ \sum_{\eta({\bf x})} $ runs over all different
field configurations  $ \eta({\bf x}) $ defined by (\ref{260}).
The values for $ v_{\rm \triangle}({\bf x}) $ can be calculated via
the Fourier transform
 \begin{equation}
  v_{\triangle}({\bf x})   =
 \frac{1}{A_{\rm BZ}} \int_{\rm BZ} \!   d^2k
 \frac{ \exp(i {\bf k} {\bf x})}{\left[\frac{4 a^2 }{9} \overline{K}_{(l)}
 K_{(m)}
 {\bf e}_{(l)}\cdot
 {\bf e}_{(m)} \right]^2}  .    \label{271}
   \end{equation}

We fix the gauge freedom due to overcounting by
setting
$ n_{l3}=0 $. Upon further using
(\ref{140}) we can express
 $ \eta({\bf x}) $ in terms of the
fields $ n_{(ij)} $ with $ i,j=1,2$ only. The
explicite relations are given in Eqs.~(\ref{a10}) and (\ref{a20}).
Figure~2(a) illustrates the non-zero $  \eta({\bf x}) $
values for $ n_{(ij)}({\bf x}) =  \delta_{i,1}
\delta_{j,2}\, \delta_{{\bf x},{\bf x}_0} $, while 2(b)
does the same thing for
$ n_{(ij)}({\bf x})=\delta_{i,1} \, \delta_{j,1} \,
\delta_{{\bf x},{\bf x}_0} $.
The cases $ n_{(ij)}({\bf x}) =\delta_{i,2} \,  \delta_{j,1}
\, \delta_{{\bf x},{\bf x}_0} $ and
$ n_{(ij)}({\bf x}) =
\, \delta_{i,2} ~\delta_{j,2} \, \delta_{{\bf x},{\bf x}_0} $
follow from linear combinations of 2(a) and 2(b)
after a clockwise rotation
by an angle $2\pi/3$. The space
of all  $ \eta$-fields are spanned by  linear
combinations of these four configurations modulo translations.
The figure shows that
the values of the $ \eta$-field are multiples
of $ 2/3 $ which we call {\em defect charges\/}. All
$ \eta $-configurations are composed
of two neutral rhombuses which are illustrated
by the dashed and straight lines in Fig~2.
Besides being neutral, the rhombuses have also no dipole moments.
The figure shows the multiplicity of the basic charges for both
rhombuses. The fact that
the $ \eta $-fields are neutral and dipole-free
agrees with the situation
on the square lattice.
There is one difference, however.
On the triangular lattice
there are neutral dipole-free charge configurations
composed of multiples of basic charges $ 2/3 $ which cannot be obtained from
 an $ \eta$-field. One simple example is a single basic rhombus shown
by the dashed line in Fig.~2(a) built
of defect charges $ \pm 2/3 $.
  \mn{example??}
The complication
with respect to
the square lattice where the $ \eta $-fields are built of all
neutral and dipole-free charge configurations
is due to the constraint (\ref{140}).
Finally we mention that the basic configuration
$ \tilde{n}({\bf x})=(2/\sqrt{3})\,  \delta_{{\bf x},{\bf x}_0} $ in (\ref{252}) with
$ \eta({\bf x})=-a^2 \nabla_{x}\nabla_{y}\tilde{n}({\bf x}) $
for $ {\bf x} \notin {\cal B} $ corresponds to
the defect configuration 2(a).

We shall  now  prove that all localized  $  \eta({\bf x})$ fields can be
built from localized $ n_{(ij)}({\bf x}) $-fields if the sample
is infinite.
%which is the main advantage of the
%$ \eta({\bf x})$  representation of (\ref{260}), (\ref{270})
% in comparison to the representation (\ref{250}) and (\ref{252}).
This can be seen from the fact that the line $ \cal B $
in (\ref{250}) and (\ref{252})
can be moved also to the middle of the system where we held fixed
the boundary $ {\cal B}' $ in (\ref{185}). This follows immediately
from the considerations in Appendix A and
the periodic boundary
condition of the displacement fields $ u_i({\bf x})$ used here.
Thus, we have to take into account only a linear
combination of the most localized fields $ n_{(ij)}({\bf x}) $
in (\ref{260}) to get the leading order in the defect partition function
(\ref{270}).

Writing  $ Z_{\triangle} $
as an exponential of the connected diagrams
$ Z=e^{- \beta  F}$,
we find the free energy of the defect fields
\begin{equation}
-F_{\triangle}^{\rm def}\! /k_B T
\! \approx \! 2N e^{-4 \pi^2 \!
\tilde{v}_{\triangle}(0) \beta_{\triangle}(1+\nu)}\!+ 6N e^{-4 \pi^2 \!
\tilde{v}_{\triangle}(0) \beta_{\triangle} \! \frac{4(1+\nu)}{3}}
 \label{272}
\end{equation}
where $ \nabla_{x} $, $ \nabla_{y} $ are the triangular lattice derivates
in (\ref{50}) and
\begin{equation}
\tilde{v}_{\triangle}({\bf x}-{\bf x})=
a^4 \nabla_{x} \nabla_{y} \nabla_{x'} \nabla_{y'} \,
v_{\triangle}({\bf x}-{\bf x}') \,.
           \label{275}
\end{equation}
By carrying out a numerical integration of the
Fourier transform similar to (\ref{205}) we obtain
\begin{equation}
\tilde{v}_{\triangle}(0)
\approx 0.163  \quad , \quad  \tilde{v}_{\triangle}(a {\bf e }_{(i)})
\approx -0.03   \,.
           \label{278}
\end{equation}
These numbers are found
by numerical integration
of the Fourier representation similar to (\ref{271}). The first term in
(\ref{272}) corresponds to the $ \eta(\bf{x}) $-configuration in Fig.~3.
The second term in (\ref{272})
corresponds to the $ \eta$-configuration 2(a) in Fig.~2. The factor
$ 6N $ comes from the six possibilities to cover this basic
$ \eta $-configuration on the lattice, this means the configuration 2(a),
the rotated configuration by angle $ 2 \pi/3 $ and
$ 4\pi/3 $ plus the negative of all these three configurations.
The other basic diagrams are approximately a
factor $ \exp[-4 \pi^2 v_{\triangle}(0)
\beta_{\triangle} (1+\nu)] $ smaller
than the leading terms in (\ref{272}). When using  the mean-field value
$  \beta_{\triangle} (1+\nu) \approx 0.6 $ we obtain additive corrections
which approximatively a factor $0.04 $ smaller than the leading terms
in $ - F_{\triangle}^{\rm def}/k_B T  $ calculated above.
From this, we derive also a temperature regime
for which our lowest defect configuration result can be trusted. This is
given by the inequality  $  4 \pi^2 v_{\triangle}(0)
\beta_{\triangle} (1+\nu) > 1 $ resulting in
$ \beta_{\triangle}(1+\nu) > 0.15$.

Recall that for square lattices, the defect correction factor
looks similar to (\ref{272}), except that the second term is missing,
and that
$ \tilde{v}_{\rm \triangle}$
in the first term is replaced by
 $\tilde{v}^{\rm \square}  $, where
$ \tilde{v}^{\rm \square} $ is defined by Eq.~(\ref{275}),
with $ v^{\rm \square} $ being
the inverse square of the
Euclidean lattice Laplacian \cite{GFCM2}.  The numerical value is
$ \tilde{v}^{\rm \square}(0) \approx 0.16 $
\cite{GFCM2}.

Next, we consider the higher-order correction factor $ Z^{\rm stress} $
to the high-temperature expansion of the partition function
$ Z=Z_{T \to \infty} Z^{\rm stress} $.
The correction factor
is obtained  by carrying out the
integration over the displacement field $ {\bf u}({\bf x}) $
 in Eq.~(\ref{190}), which makes $ \sigma_{12} ({\bf x})$ discrete-valued,
and further
summing over the defect gauge fields $ n_{(ij)}({\bf x}) $. The result is
\begin{align}
& Z_{\triangle}^{\rm stress}  =  \prod_{{\bf x}}
\left((1-\delta_{{\bf x},{\cal B}'})
\left[\sum_{\sigma_{12}({\bf x}) \in \sqrt{3}
{ \mathbb Z }/2}\right] \right.
 \nonumber \\
&
\left.+\delta_{{\bf x},{\cal B}'}
  \left[\sum_{(\frac{1}{\sqrt{3}}-
\frac{\overline{\nabla}_x}{\overline{\nabla}_y}) \sigma_{12}({\bf x}) \in
   \mathbb Z }\right]
\left[\sum_{(\frac{1}{\sqrt{3}}-\frac{1}{2}
\frac{\overline{\nabla}_y}{\overline{\nabla}_x}+\frac{1}{2}
\frac{\overline{\nabla}_x}{\overline{\nabla}_y})\sigma_{12}({\bf x}) \in
  \mathbb Z }\right]\right)  \nonumber \\
&
 \times\exp\Bigg[
\frac{-1}{4 \beta_{\triangle}(1+\nu)} \sum_{{\bf x}, {\bf x}'}
\sigma_{12}({\bf x}) \Bigg[\frac{4a^2}{9} \,
\overline{\nabla}_{(l)}{\bf e}_{(l)}{}_x
\overline{\nabla}_{(m)}{\bf e}_{(m)}{}_y \,
\nonumber \\
& ~~~~\times v_{\rm \triangle}({\bf x}-{\bf x}')
\; \frac{4a^2}{9}\nabla_{(k)}{\bf e}_{(k)}{}_{x'}
\nabla_{(n)}{\bf e}_{(n)}{}_{y'}
\bigg]^{-2} \sigma_{12}({\bf x}') \Bigg]
 .  \label{280}
\end{align}
Most of the $ \sigma_{12} $ configurations give a zero contribution
to $ Z_{\triangle}^{\rm stress} $ in the sum
$\sum_{\sigma_{12}({\bf x}) \in \sqrt{3} { \mathbb Z }/2} $ in
(\ref{280}). We can extract the non-zero contributions by going to
the gauge field $ \chi({\bf x}) $ with the definition
$ \sigma_{12}({\bf x})= a^2 \nabla_x \nabla_y \chi({\bf x}) $.
In order to get the summation values of the fields $ \chi({\bf x})$
we have to determine the lowest value  $ z $  such that
$  z  a^2 \nabla_x \nabla_y \delta_{{\bf x}, {\bf x}_0}
\in \sqrt{3} {\mathbb Z}/2$
for an arbitrary lattice position $ {\bf x}_0 $. This value is given
by $ z=3/2 $.
Neglecting boundary terms, we
obtain
\begin{align}
& Z_{\triangle}^{\rm stress}  =  \prod_{{\bf x}}  \left[
\sum_{\chi({\bf x}) \in \, 3 \,{ \mathbb Z }/2}    \right]  \,.
      \label{300}
      \\
&  \times \exp\left[-\frac{1}{4 \beta_{\triangle}}
\frac{1}{(1+\nu)} \sum_{{\bf x},{\bf x}'} \chi({\bf x})
(v_{\triangle})^{-2} ({\bf x}-{\bf x}')
 \chi({\bf x}') \right]   \nonumber
\end{align}
where
$v_\triangle^{-2}  ({\bf x}) $ is short for
$ a^4\, \square^{2}_{\triangle} ({\bf x})$ [recall (\ref{255})].
Numerical integration yields
\begin{equation}
 (v_{\triangle})^{-2}(0)\approx 18.5 \quad      ,  \quad
(v_{\triangle})^{-2}(a {\bf e}_{(i)})
 \approx -4.43  \,,                   \label{315}
\end{equation}
so that the lowest correction to the free energy is
\begin{equation}
  -F_{\triangle}^{\rm stress}/k_BT 
\approx 2 N e^{-{9 (v_{\triangle})^{-2}(0)}/{16
 \beta_{\triangle}(1+\nu)}} \,.            \label{320}
\end{equation}
The next term in the expansion of $  -\beta F_{\triangle}^{\rm stress} $
is about a factor  $ \exp[-9 (v_{\triangle})^{-2}(0)/(16
 \beta_{\triangle}(1+\nu))] $ smaller than the lowest-order term in
(\ref{320}).
By taking into account the lowest-order result
$ \beta_{\triangle}(1+\nu) \approx 0.6 $ we obtain from
this an extremely small factor $ e^{-20} $. Thus, the lowest
stress configuration of the high-temperature expansion in (\ref{320}) can be
trusted for $ \beta_{\triangle}(1+\nu) < 10 $.

In the case of the square lattice \cite{GFCM2}  one gets for
$ Z_{\square}^{\rm stress} $ the same formula as in (\ref{320}), except that
the range of summation for the gauge field $\chi({\bf x})$ is
$ \sum_{\chi({\bf x}) \in \, { \mathbb Z }} $, and that
$ (v_{\triangle})^{-2}$ has to be
replaced by $ (v_{\square})^{-2} $. 
The numerical values are
 $ (v_{\square})^{-2}(0)=20 $ and
$ (v_{\square})^{-2}(a {\bf e}_{x,y})=-8$.

Using the above-derived results we can  now
calculate the melting temperature
from the intersection of the
low-temperature expansion $ Z_{T \to  0} Z_{\triangle}^{\rm def} $ and
the high-temperature expansion $ Z_{T \to  \infty} Z_{\triangle}^{\rm stress} $
of the partition function $ Z $. Instead of
 (\ref{240}), we find from the corrections
(\ref{272}) and (\ref{320}) that the melting point satisfies
\begin{eqnarray}
 \beta  (1+\nu) & \approx &  A   \exp\bigg[- M_1 e^{-B_1
\beta (1+\nu)} -  M_2 e^{-B_2
\beta (1+\nu)}  \bigg]   \nonumber \\
& &\times  \exp\bigg[+N  e^{- \frac{C}{\beta (1+\nu)}} \bigg]
               \label{330}
\end{eqnarray}
with  $ \beta= \beta_{\triangle} $ in the case of the triangular
lattice and  $ \beta= \beta_{\square} $ for the square lattice.
The parameters are
\begin{equation}
\begin{array}{c @{\hspace{0.3cm}} |@{\hspace{1cm}} c @{\hspace{1cm}} c }
 & {\rm triangular} & {\rm square}  \\
  \hline
 A   & 0.51 & 0.81  \\
  M_1   & 4 & 4  \\
 M_2   & 12 & -  \\
 N   & 4 & 4  \\
 B_1  &  6.45 & 6.31  \\
 B_2  &  8.6  & -  \\
 C   &  10.4  & 5  \\
\hline
\end{array} \label{340}
\end{equation}

Expression
$ \tilde{ F}/N k_B T \equiv  
F/N k_B T + \ln((1-\nu)^{1/2} (1+\nu))/N $
is a function of
$ \beta(1+\nu) $.
We show in Fig.~ 4 the corresponding curves
of the low- and high-temperature expansion for the triangular
lattice and the square lattice.
We obtain for the square lattice that the high and low-temperature curves
intersect in one point.
In contrast to this, the curves of the triangular
lattice do not intersect. This is caused by
an enhancement
of the defect contributions to the low-temperature expansions for the
triangular lattice. The two curves have merely a would-be
intersection  near
\begin{equation}
\beta_{\triangle} (1+\nu) \approx  0.6,
 \label{@betatr}\end{equation}
where the
distance between both curves
on the
$ \beta (1+\nu) $-axis
 is approximatively
\begin{equation}
\Delta \beta_{\triangle} (1+\nu) \approx 0.05 .
\label{@betatr2}\end{equation}
We point out that the non-intersecting of the two curves
is not reasoned in a failure of our approximations. As estimated
above the two curves can be
trusted for $ 0.15 <\beta_{\triangle}(1+\nu) <10 $.

For a square lattice there was a definite intersection at
\begin{equation}
 \beta_{\square} (1+\nu)   \approx  0.8 .
\label{@betasq}
\end{equation}
In either case,
the
higher-order corrections to
the melting temperature are small
compared to
the lowest-order result (\ref{240}).
What is the
reasons  for the difference in the transition properties of the
two lattices?
We see from (\ref{340}) that the numbers
$ M_1 $ and $ B_1 $ are almost identical
for the triangular and the square lattice. The
$ \eta $-configuration for the triangular lattice
of this term is a triangle with charges
$ \pm 2/3 $ shown in Fig.~3.
The corresponding $ \eta $-configuration for the square lattice
is given by a basic square with charges $ \pm 1 $ \cite{GFCM2}.
The difference between
 triangular and the square lattices
lies in the term proportional to $ M_2 $
in   (\ref{330})
whose $ \eta $-configuration
is shown in
Fig.~2(a).
A  term of this type
is absent
for square lattices.
The defects lie on two neighboring rhombuses of charges $ \pm 2/3 $.
The corresponding configuration for the square lattice
would consist
of two   neighboring  squares.
The distance
between the rhombuses
is much
smaller than that of
the two nearby squares.
This results in an energy reduction of
the corresponding defect configuration $ \eta $ for the
triangular lattice. This was the reason why
this contribution was negligible for square lattices.

  \begin{figure}
% \hspace*{0.1cm}
\begin{center}
\psfrag{betaF/N}{\scriptsize $ \tilde{F}/N k_B T $}
\psfrag{beta(1+nu)}{\scriptsize $ \beta(1+\nu) $}
\psfrag{low beta}{\scriptsize low $ \beta $}
\psfrag{high beta}{\scriptsize high $ \beta $}

\includegraphics[height=7cm,width=7.8cm]{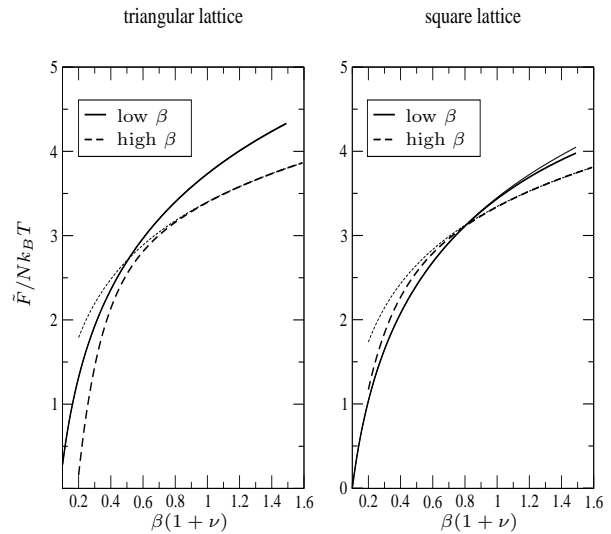}
\end{center}
\hspace*{-0.2cm}
\caption{Low- and high-temperature expansions of
$\tilde{F}/Nk_B T \equiv F/N k_B T + \ln((1-\nu)^{1/2} (1+\nu))/N $
for the triangular lattice and the square lattice. The thin line curves
are the lowest-order results of Section V.A, while the
thick line curves take into account also
the defect and stress contributions of Section V.B.
Note that stress corrections are negligible. The low-temperature
lowest-order and higher-order curves for the triangular lattice
are practically on top of each other.}
\end{figure}

From previous work we know
that melting of two dimensional crystals can proceed
in only
two possible scenarios: either by the proliferation of defects
as in three dimensions, in which case the transition is of first order,
or
by a two-step unbinding of defects (KTHNY). It was
shown by Saito \cite{saito1} in  Monte Carlo simulations
of a gas of dislocations on a triangular lattice
and in the textbook \cite{GFCM2} by general arguments
that
the core energies of the dislocations
decide which of the two options is chosen.
For small core energies, the
transition is of first order,
for large core energies,
the KTHNY scenario prevails.
This agrees with the properties of our harmonic
defect model.
From the long-range
behavior of the Coulomb force between the
dislocations
the unbinding transition will always take place
even if the defects are not yet prolific at a temperature given by the universal
equation KT criterion. This implies that the $ \beta $-value at the
melting point must  always satisfy the inequality
$ \beta_{\triangle}(1+\nu) \ge \sqrt{3}/\pi \approx 0.55  $
irrespective of the lattice symmetry \cite{halperin1}.
If the inequality is almost an equality,
which is the case for
our result (\ref{@betatr}) ,
the transition is of
the KTHNY type.
For square lattices, the situation is different.
By comparing the melting temperature of the square (\ref{@betasq})
and triangular (\ref{@betatr}) lattice we see
that defects proliferate {\em before\/} dislocations can separate.
This agrees
with the naive expectation that square lattices
should be less stable against thermal fluctuations than triangular ones.
In the lattice
models this is a consequence of the
smaller
core energies of dislocations on square lattices
\cite{GFCM2}, as we know from
Saito's simulation work.

\comment{
Our findings agree with
the Monte Carlo simulations carried out by Saito \cite{Saito1}
for a gas of dislocations on a triangular lattice.
These show
a strong dependence of the nature
of the transition (first-order versus KTHNY-like),
on the core energies of the dislocations.
If the core energy of the
dislocations exceeds a certain value, the dislocations first unbind
before they proliferate, resulting in a KTHNY transition. \mn{war das nur eine transition?}
The difference between
 triangular and square lattices in our model
is of a similar origin.
%%%%%%%%%%%%%%%%%
Due to the enhancement
of the number of defect configurations of lowest energy on
the triangular lattice,
the relative width of the atomic displacements
is larger than on a square lattice.
By Lindemann's rule \cite{lindemann1},
the melting point depends
mainly on this quantity.
In two dimensions, this leads to an earlier onset
of dislocation unbinding transition.}

The behavior of the
free energy
shown in Fig.~4
{\em excludes\/} the
single-first-oder scenario for triangular lattices, leaving only
 the KTHNY scenario.
Indeed, for this scenario we cannot expect
an intersection of the
curves in Fig.~4.
The intermediate hexatic phase between
the two continuous transitions, which the model
should exhibit,
would be inaccessible to both low- and  high-temperature expansions.
With less certainty, we extract an estimate
for the separation of the
two KT transitions
from the
 distance $ \Delta \beta $
of the two nonintersecting curves. 
Our finding of a KTHNY scenario
is in accordance with recent computer simulations
for triangular Lennard-Jones  as well as
electron lattices
\cite{udink1,chen1,muto1,he1}.

The situation for the
square lattice
is much clearer.
There is a definite
intersection point in Fig.~4, indicating that the melting
transition is of first order, as found by computer simulations
 \cite{kleinert86,GFCM2}.
When trying to compare
this result with real crystals
in nature one has the difficulty that square lattices are hard to produce.
They need complicated interparticle forces, and only
Weber et al. \cite{weber1} succeeded to do
build such crystals
with the help of a  three-body potential in a computer model.
They found indeed a first-order melting transition
in their simulation,
in agreement with the
prediction in  \cite{kleinert86,GFCM2}, and with the above conclusion.

\section{Lennard-Jones and electron lattices in 2D}
Let us compare our results quantitatively
with computer simulations of  2D Lennard-Jones
as well as electron  lattices. In  both cases,
the ground state is triangular, due to
the simplicity of the interaction potential.
In the Lennard-Jones lattice,
the interaction potential is
\begin{equation}
V_{LJ}(r)=4 \epsilon\left[\left(\frac{\sigma}{r}\right)^{12}-
\left(\frac{\sigma}{r}\right)^{6} \right] \,.          \label{500}
\end{equation}
Frenkel and McTargue \cite{frenkel1} carried out an isothermal-isochoric
molecular dynamics simulation, and observed a hexatic phase
in accordance with KTHNY theory. In contrast, Abraham at. al.
and others \cite{abraham2,koch1} found
with the help of
both molecular dynamics
as well as
 Monte-Carlo  simulations
 that the melting transition
of the Lennard-Jones lattice is
of first order.
The discrepancy induced
 simulations on larger systems, which found again
KTHNY-like
melting transitions \cite{udink1,chen1}.

To apply our model,
we extract from the potential (\ref{500})
the elastic moduli $ \lambda $ and $ \mu $ as follows \cite{GFCM2}:
\begin{eqnarray}
\lambda & = &   \frac{\epsilon}{v}  \,
\left[324 \left(\frac{\sigma}{a}\right)^{12}  -
108  \left(\frac{\sigma}{a}\right)^{6}  \right] \,,  \label{510} \\
\mu & = &   \frac{\epsilon}{v}  \, \left[180
\left(\frac{\sigma}{a}\right)^{12}
-36 \left(\frac{\sigma}{a}\right)^{6}   \right] \,.  \label{520}
\end{eqnarray}
Inserting these into our result (\ref{@betatr}),
we obtain for the melting temperature $ T_m $
of
the triangular Lennard-Jones
lattice:
\begin{eqnarray}
\frac{k_B T_m}{\epsilon} & \approx & 112 \,
(\sigma^2 \rho)^3 [(\sigma^2 \rho)^3 -0.3] \frac{1}{0.6 \, (2 \pi)^2}
\label{525}            \\
 & \approx & 4.725
\, (\sigma^2 \rho)^3 [(\sigma^2 \rho)^3 -0.3] \,.   \nonumber
 \end{eqnarray}
Here $ \rho $ is the particle density $ 1/ v $.
Figure~5 shows $ k_B T_m / \epsilon $ as a function
of $ \sigma^2 \rho $ in comparison with the
 two melting curves
 enclosing the coexisting crystal-liquid region or the hexatic phase,
respectively, calculated by
 Abraham and Barker et al. in \cite{abraham2} getting a first-order
transition and by Udink et al. in
\cite{udink1} consistent with KTHNY.
 The curves are  calculated by determining the phase transition line via
thermodynamic integration. The results in \cite{udink1}
are in good accordance with the simulations on large systems showing also
a KTHNY transition.

Obviously,
the value $  0.6 $  in
relation (\ref{@betatr})
is too small to agree with the
simulations. This discrepancy will be
removed below in Section~VII.
 \\

\begin{figure}
\begin{center}
\psfrag{rho sigma^2}{\scriptsize{$\rho \sigma^2$}}
\psfrag{kT/epsilon}{\scriptsize{$k_BT/\epsilon$}}
\includegraphics[width=8cm]{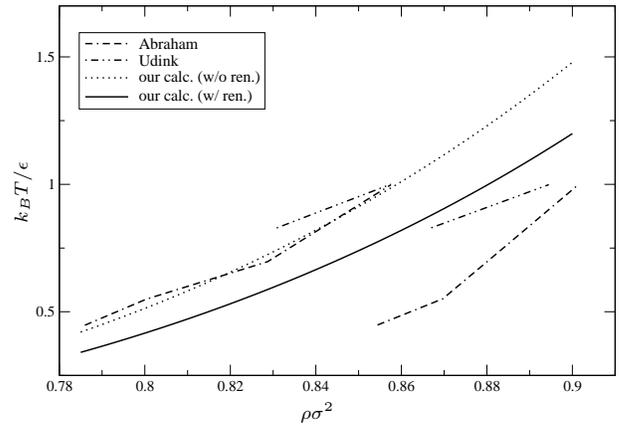}
\end{center}
\caption{
Reduced melting temperature $ k_BT_m/\epsilon $
of Eq.~(\ref{525})
 as a function of the reduced density $ \rho \sigma^2  $.
The dotted
line denotes the melting temperature calculated with the
$T=0 $ elastic constants.
The solid curve takes the thermal softening of the
elastic constants
by non-harmonic elasticity into account,
calculated in Section~VII.
For comparison we show
the melting temperature obtained by
Abraham and Barker et al. \cite{abraham2} and Udink et al.
\cite{udink1} by  simulation.
The two curves of the simulation enclose the coexisting
crystal-liquid region or the hexatic phase, respectively.
\comment{
Reduced melting temperature $ k_BT_m/\epsilon $
of the 2D Lennard-Jones lattice
 as a function of the rescaled density $ \rho \sigma^2  $
according to our prediction
(\ref{525}).
 The dotted
line is obtained
by inserting
the
elastic constants of the $T=0$ -crystal
into our
formula  (\ref{@betatr}),
the solid curve takes their thermal softening
due to
the anharmonic elastic terms into account, calculated
in Section~V.  It is compatible
with the
solid curves
enclose the coexisting
crystal-liquid region
in the simulations of
Abraham and Barker et al. \cite{abraham2}.}
}
\end{figure}

Next, we discuss the system of a 2D electron lattice.
Wigner \cite{wigner1} predicted in 1934 that a gas of electrons should
becomes
a triangular crystal at
low temperature and density.
Experimentally, this was first
observed by Grimes and Adams \cite{grimes1}
for electrons trapped on a
liquid helium surface.
The electrons interact via
the Coulomb potential
\begin{equation}
V_{ee}(r)= { e^2}/{r}  \,,           \label{530}
\end{equation}
and their kinetic energy is
negligible at low temperature and density.
As in the case of the Lennard-Jones lattice, the ground state of the system
is built of a triangular lattice \cite{bonsall1}.
Hockney and Brown \cite{hockney1} were the first which
carry out a  molecular dynamic simulation  of this system.
They found a $ \lambda $-like melting transition.
In contrast, Gann et al. and Kalia et al. \cite{gann1} found,
 via Monte-Carlo and
molecular dynamics studies,
a first-order melting transition. Differing from both results,
Morf \cite{morf1} obtained
a melting temperature
consistent with the
KTHNY theory,
when he took into account the higher-order
nonlinear corrections to the shear modulus.
His results were reproduced
by more recent simulations on
larger electron lattices
 \cite{muto1,he1}.

The elastic constants for the electron lattice are
calculated in Refs. \cite{bonsall1,GFCM2}.
The  Lam\'e constant is infinite,
so that the
electron lattice has an infinite modulus of compression.
 The
 shear constant $ \mu $ is equal to
\begin{equation}
\mu= \eta e^2/v^{3/2},~~~~
 \eta\approx0.245 .
\label{@MU}\end{equation}
The simulation data yield for the melting temperature $ T_m $
the relation
\begin{equation}
 \Gamma_m   \approx 130 \pm 10 ,
 \label{@SV}\end{equation}
where $ \Gamma $ is the reduced
inverse temperature
\begin{equation}
 \Gamma \equiv \sqrt{\pi} e^2/\sqrt{v} k_B T.
\label{@GAM}\end{equation}
To compare this result with our
theory
we insert
the elastic constants
into Eq.~(\ref{@betatr}) and find
$ \Gamma_m \approx 86.4 $, which is about 30\% smaller than the simulation value.
Also here, we shall remove the discrepancy
in Section~VII.

Concluding this section we observe that
the interval
$ \Delta \beta_{\triangle}(1+\nu) \approx 0.05 $
found in Fig.~4 for the triangular lattice
in Eq.~(\ref{@betatr2}) agrees
with the temperature interval of
the intermediate (hexatic) phase
  seen in the
computer simulations of both Lennard-Jones as well as
electron lattices \cite{udink1,chen1,muto1,he1}.

\section{Anharmonic Corrections to Melting Formula}

The melting temperature
found in the last section
from formula (\ref{@betatr}) lies too high.\mn{not too low as you write}
The same problem was encountered by
Thouless \cite{thouless1}
in his calculation of the melting temperature for the electron lattice,
who used
 the melting formula $ \beta_{\triangle}(1+\nu) =\sqrt{3}/\pi \approx 0.55 $
where dislocations unbind in the
KTHNY theory.
%(in contrast to our melting formulas in (\ref{@betatr}), (\ref{@betatr2}) the
%elasticity constants $ \mu $ and
%$ \lambda $ in this melting formula
%contains also corrections due to the dislocations and disclinations).
His formula  is quite close to
ours in Eq.~(\ref{@betatr}). It was shown
by Morf in \cite{morf1} that the discrepancy
between the prediction
with the simulated melting temperature of the electron lattice
can be explained
by the thermal softening
of the elastic constant $\mu$, which he
 determined with the help of a computer simulation to be [recall
(\ref{@GAM})]
\begin{equation}
\mu(T)\approx \mu(0)(1-30.8 /\Gamma) \,,
                   \label{540}
 \end{equation}
where $ \mu(0)  $
is the zero-temperature shear modulus
(\ref{@MU}).
By using (\ref{540}) one obtains from our formula (\ref{@betatr})
$ \Gamma_m \approx 117.2 $, which is now close to Morf's simulation value
(\ref{@SV}) and in quite excellent agreement with the most recent
melting value in  Ref.~\onlinecite{he1} obtained by Monte-Carlo simulation
of  a large Wigner lattice $ \Gamma_m \approx 123 $.

For 2D Lennard-Jones crystals,
the thermal softening
can only be estimated theoretically.
\comment{The renormalization of the Lennard-Jones
elastic constants due to non-linear
elasticity effects has not been calculated in general.}
\mn{ist doch in meinem Buch, oder?
Muss bekannt sein fuer phonon Leute. Zimann "Electrons and Phonons"}
We obtain a reasonable
approximation of the renormalized elastic
constants by
observing
that the thermal softening
due to non linear elasticity
is much stronger for the transversal
than for the longitudinal sound velocity \cite{GFCM2}.
A further simplification
comes from the fact
 \cite{GFCM2}
that the renormalization of the transversal sound is proportional
to the square root of $ \mu $, and has in first approximation
a universal low-temperature law depending only on the lattice
structure. This yields the approximations
\begin{align}
& 2 \mu(T) +\lambda(T) \approx   2 \mu(0) +\lambda(0) \,,  \label{555} \\
& \mu(T)   \approx \mu(0)\left[1- c \, \frac{\mu(0)}{\beta(0)}
\left(\frac{1}{\mu(0)}+ \frac{1}{2 \mu(0)+\lambda(0)} \right)\right] \,.
\label{558}
\end{align}
%Here $ \mu(T) $, $ \lambda(T) $, $ \nu(T) $, and $ \beta(T) $ are
%given by the temperature renormalized corresponding values.
The constant $ c $ is the same for all lattices with the same structure.
Its value can therefore be deduced
 from (\ref{540}):
\begin{equation}
c_{\triangle} \approx  0.11 \;. \label{560}
\end{equation}
Combining (\ref{555}) and (\ref{558}) we obtain
%for  low temperatures, where  $ c / \beta(0) \ll 1$,
\begin{eqnarray}
\mu(T) & \approx & \mu(0)
\left[1- \frac{c}{\beta(0)}\left(\frac{3}{2}-\frac{\nu(0)}{2}\right)\right]
 \,,
\label{570}  \\
\nu(T) & \approx & \nu(0)+(1-\nu(0))
\frac{c}{\beta(0)}\left(\frac{3}{2}-\frac{\nu(0)}{2}\right) \,.\label{575}
\end{eqnarray}
Taking
this renormalization
into account, we obtain
from (\ref{@betatr}) the melting formula for triangular lattices
\begin{eqnarray}
 \beta_{\triangle}(T)(1+\nu(T))
& \approx &\beta_{\triangle}(0)(1+\nu(0))
-c \;\nu(0)(3-\nu(0))\nonumber \\& \approx& 0.6 \,. \label{580}
\end{eqnarray}
For the Lennard-Jones lattice where $\nu(0) \approx 0.48$, this leads
to a melting formula
\begin{equation}
\beta_{\triangle}(0)(1+\nu(0)) \approx  0.74,
\label{@}\end{equation}
rather than (\ref{@betatr}), thus lowering
the melting temperature by about $20\%$.
The dashed curve in Fig.~5
shows the resulting melting temperature
for the Lennard-Jones lattice, which is compatible with the
simulation data
 \cite{udink1,abraham2}.
It is obtained from
(\ref{525}) by multiplying the right-hand side with
the softening factor
$ 0.6/0.74 \approx 0.81 $.

\section{Two-dimensional Lindemann parameter of melting}
The Lindemann parameter \cite{lindemann1}
expressed in terms of
the expectation value of the lattice displacement
$ \sqrt {\langle u^2({\bf r}_i) \rangle}/a $  is not applicable in
two dimensions, where the expectation value
diverges due to excessive long-wavelength fluctuations of the
displacement field.
An appropriate modified Lindemann
parameter in 2D is  \cite{bedanov1,zahn1,zahn2}
\begin{equation}
\tilde{L} ={ \sqrt{ \langle \left[u({\bf x}_i)-u({\bf x}_{i+1})\right]^2
\rangle }}/{a}    \,,             \label{600}
\end{equation}
where $ {\bf x}_i $ and $ {\bf x}_{i+1}$  are nearest neighbors in the
lattice.
The
correlation function of the
displacement field
following from
the energy (\ref{85}) is
\begin{align}
& \langle u_i({\bf k}) u_j(-{\bf k}) \rangle =  N (k_B T)\label{610} \\
&
\times \frac{
\overline{K}_{(l)} K_{(m)} {\bf
e}_{(l)}{\bf e}_{(m)}
\, \delta_{ij} -
\frac{\lambda + \mu}{\lambda + 2 \mu}
 \overline{K}_{(l)} K_{(m)}  {e_{(l)}}_i{e_{(m)}}_j
}
{ v \mu \,  \frac{4}{9}\;(\overline{K}_{(l)}
K_{(m)}  {\bf e}_{(l)}{\bf e}_{(m)})^2 }  .
        \nonumber
\end{align}
On square lattices, we must merely replace
$ (2/3) K_{(l)} ({e}_{(l)})_{x,y} \rightarrow
K_{x,y}$ and
$ (2/3)\overline{K}_{(l)}
({e}_{(l)})_{x,y} \rightarrow
\overline{K}_{x,y} $ \cite{GFCM2}.
From this equation
 we obtain
for both  square and
triangular lattices the modified  Lindemann parameter
\begin{equation}
\tilde{L}=
\frac{1}{4 \pi} \sqrt{\frac{3 - \nu(T)}{\beta(T)}}  \,.
\label{620}
\end{equation}

Let us compare our Lindemann parameter
with that determined by
 Bedanov and Gadiyak via computer simulations
for Lennard-Jones and
electron lattices
\cite{bedanov1}.
For  the electron lattice we insert
the
temperature-dependent elastic constants $ \mu(T) $ of Eq.~(\ref{570})
and $ \nu(T)=\nu(0)=1 $ into
the melting formula  (\ref{580}),
and
 obtain from (\ref{620})
$ \tilde{L}_{\triangle} \approx  0.20$.
The corresponding simulation value of
Bedanov and Gadiyak is $ \tilde{L}_{\triangle} \approx  0.17 $, quite close to this.

Now we turn to the
Lennard-Jones lattice.
Using the renormalized
elastic constant (\ref{575}) with $ \nu(0) \approx 0.48 $
and the melting formula  (\ref{580}) we obtain
from
(\ref{620})
the  Lindemann parameter $ \tilde{L}_{\triangle} \approx  0.20 $, again very close to
the value
$ \tilde{L}_{\triangle} \approx 0.18 $
found in the simulations of
Bedanov and Gadiyak \cite{bedanov1}.

Hence our theory yields
generalized
Lindemann parameters for
 both electron
and Lennard-Jones lattices in good agreement with
simulation results
of Bedanov and Gadiyak \cite{bedanov1}.

\section{Summary}

We have set up the simplest possible
lattice model of defect melting
on a two-dimensional
triangular lattice.
It accounts for the correct elastic fluctuations
and by means of
discrete-valued defect gauge fields
for the fluctuations of dislocations and disclinations.
The latter give rise to melting transitions, and
the melting temperature follows a formula
of the Lindemann type, with a modification
due to the two dimensions.
The value of the Lindemann parameter is predicted, and
 agrees with estimates obtained from computer simulations
on both Lennard-Jones and electron lattices.

The melting transition is determined from the
intersection of the free energies calculated once from a
low-temperature expansion and once from a
 high-temperature expansion.
While the square lattice
melts in a first-order transition,
the
curves for the
triangular lattice are compatible with two continuous transitions,
as predicted by the KTHNY theory and
found in computer
simulations.
The difference is due
to the enhancement of the defect fluctuations
in comparison to the square lattice
caused
by the smaller distance
between basic nearby rhombuses on  triangular  lattices,
 in comparison
to the corresponding defect configuration
of two nearby squares on the square lattice.
The resulting
enhancement of smallest defect configurations leads
to a decrease
 of the melting temperature.

\comment{
Next, we compare our melting results applied to  the Lennard-Jones
system as well as the electron lattice with
the corresponding computer simulations.
 When taking into account the temperature dependence of the
elastic constants which are due to nonlinear elastic effects
we obtain good accordance to the melting temperatures
obtained by computer simulations.
We calculate  Lindemann like numbers for the 2D
square and triangular lattice. These are given by the square root expectation
value of the next nearest neighbors distance.
When comparing the calculated Lindemann numbers
for the triangular lattice
with the corresponding values gained by computer simulations we obtain
good agreement.
}
\begin{appendix}
\section{Elimination of the gauge degrees of freedom}
In this section we eliminate the gauge degrees of freedom
in the $ n_{(lm)} $ sum of Eq.~(\ref{130})
enforced by the functional $  \Phi[n_{(lm)}] $.
We shall prove the following:
The gauge degrees of freedom in (\ref{130}) with (\ref{120}) and (\ref{140})
are fixed in the case that one chooses the gauge fixed
integer-valued defect fields (\ref{180}) with the
boundary condition (\ref{185}) for the bulk defect field $n $.

By using (\ref{30}) and (\ref{140}) one can transform (\ref{120})
to the following expression
\begin{eqnarray}
 \nabla_{x} {\bf u}({\bf x}) &   \rightarrow &
 -\nabla_{(2)}  {\bf u}({\bf x})
 + n_{(2m)}({\bf x})
 {\bf e}_{(m)}
\nonumber  \\
 & = & -\nabla_{(2)}  {\bf u}+
  \left(\begin{array}{c}
  \frac{1}{2} n_{(21)}-n_{(22)} \\
  \frac{\sqrt{3}}{2} n_{(21)}
  \end{array}
\right) \,,
  \label{a10}   \\
 \nabla_{y} {\bf u}({\bf x}) &   \rightarrow &
 \frac{1}{\sqrt{3}} (2\nabla_{(1)}+ \nabla_{(2)}) {\bf u}({\bf x})
                   \nonumber         \\
 & &  - \frac{1}{\sqrt{3}} (2 n_{(1m)}{\bf e}_{(m)}+ n_{(2m)}{\bf e}_{(m)} )
 \nonumber \\
&  = & \frac{1}{\sqrt{3}} (2\nabla_{(1)}+ \nabla_{(2)}) {\bf u}({\bf x})
                   \label{a20}        \\
& & +  \frac{1}{\sqrt{3}} \left(\begin{array}{c}
 -n_{(11)}+2n_{(12)}-\frac{1}{2} n_{(21)}+n_{(22)} \\
  -\sqrt{3} n_{(11)}-\frac{\sqrt{3}}{2} n_{(21)}
  \end{array}
\right) \,.
\nonumber
\end{eqnarray}
Here we put $ n_{(l3)}=0 $ which is possible due to the
overcounting of the basis ${\bf e}_{(l)} $.
Now, we can carry out the substitution
of the displacement fields $ {\bf u}({\bf x}) \rightarrow
{\bf u}({\bf x})+a {\bf N}({\bf x}) $ in (\ref{a10})
where $ a {\bf N}({\bf x}) $ is some displacement field corresponding
to a jump from one lattice site to another.
We determine it by the requirement that
\begin{align}
&  a \nabla_{(2)}   N_x({\bf x})  =
        {n_{(2m)}{e}_{(m)}}_x
=  \frac{1}{2} n_{(21)}-n_{(22)}
   \label{a30}
\end{align}
and
\begin{align}
& \frac{a}{\sqrt{3}} (2\nabla_{(1)}+ \nabla_{(2)})  N_y({\bf x}) \nonumber \\
&  =  \frac{1}{\sqrt{3}} (2 n_{(1m)}{{e}_{(m)}}_y+
{n_{(2m)}{e}_{(m)}}_y ) =n_{(11)} +\frac{1}{2} n_{(21)}
 \,.  \label{a40}
\end{align}
From these equations we obtain
\begin{equation}
a \nabla_{(l)} N_x({\bf x}) \in \frac{{\mathbb Z}}{2}\quad \,, \quad
 a \nabla_{(l)} N_y({\bf x}) \in \frac{\sqrt{3}}{2} {\mathbb Z}
\label{a45}
\end{equation}
 and
\begin{equation}
\left(2 a \, \nabla_{(2)} N_x({\bf x})\right)\; {\rm mod} \;2 =
\left(\frac{2a }{\sqrt{3}} \nabla_{(2)}
N_y({\bf x})\right) \; {\rm mod}\;2   \,.
\label{a50}
\end{equation}
To derive the last equation we have used the fact
that the half-integer terms
on the right hand sides of (\ref{a30}) and (\ref{a40}) coincide.
We get unique solutions of (\ref{a30}) and (\ref{a40})
when fixing  $ N_x({\bf x}) $ and $ N_y({\bf x}) $ on
one half of the boundary of the system where we now suppose that we have
approximately a square sample. Thus, we can for example
fix the values of
$ N_x({\bf x}) $ and $ N_y({\bf x}) $ on the upper and the rightmost
boundary of the sample which we denote by $ {\cal B} $. Then, due
to the periodic boundary conditions for the displacement field
$ {\bf u} $ the values of $ N_x({\bf x}) $ and $ N_y({\bf x}) $
are determined on the whole boundary of the system.
Because of the periodic boundary conditions for
$ N_x({\bf x}) $ and $ N_y({\bf x}) $ it is not clear that the unique
solution of (\ref{a30}) and (\ref{a40}) respects these periodic boundary
conditions. That this is not generally true can be most easily
seen for the square lattice \cite{GFCM2}.
Whether  this is true or not depends on the values
of the defect configuration on the right hand sides
of (\ref{a30}), (\ref{a40}) and the
boundary values of $ N_x({\bf x}) $ and $ N_y({\bf x}) $ on $ {\cal B} $.
%We will consider this problem further below.

The values of the field $ n({\bf x}) $ in (\ref{180}) or (\ref{190})
are
now given by the following expression
\begin{eqnarray}
n({\bf x})& = & \frac{\sqrt{3}}{2}
 \bigg[a \left(\nabla_x N_y({\bf x})+\nabla_y N_x({\bf x})\right)
\nonumber \\
& & +n_{(2m)} {e_{(m)}}_y-\frac{1}{\sqrt{3}}(2n_{(1m)} +n_{(2m)}){e_{(m)}}_x
\bigg]    \nonumber  \\
& = & -2 n_{(11)}+n_{(12)}   \nonumber \\
& & +a \, \nabla_{(1)} \, N_x({\bf x})+a \, \sqrt{3}
\, \nabla_{(1)}
  N_y({\bf x}) \,.   \label{a60}
\end{eqnarray}
Here we take into account that the Hamiltonian (\ref{80}) depends
on the lattice derivates of the displacement
field
only in the strain combination
$ \nabla_x u_y + \nabla_y u_x $.
We now observe that $  \nabla_{(1)} N_{x,y}({\bf x}) $
can be written
as a linear combination of $ \nabla_{(2)} N_{x,y}({\bf x}') $
with $ y'=y $ or $ y'=y+{e_{(1)}}_{y} $
and a lattice derivate corresponding to a rightmost boundary edge.
From this, (\ref{a45}), (\ref{a50}),
we obtain that $ n({\bf x}) $ on the whole lattice,
given by the right-hand side of (\ref{a60}), is integer
valued when $ n({\bf x}) $ on the boundary
$ {\bf x} \in {\cal B} $ have integer
values. \mn{muss das nicht andersherum?}

The boundary values $ N_x({\bf x}) $ and
$ N_y({\bf x}) $ on $ {\cal B} $ are fixed by choosing the boundary conditions
(\ref{185}) for $n({\bf x}) $. The periodic boundary conditions for $ N_x({\bf x}) $ and
$ N_y({\bf x}) $ on ${\cal B} $ can be fixed when taking into account
(\ref{a10}) and (\ref{a20}) only for $ {\bf x} \not= {\cal B} $.
From (\ref{a10}) and (\ref{a20}) we obtain
further that we can fix  $ n_{(22)}({\bf x}) = 0 $ on  $ {\cal B} $.

As a result we
obtain the gauge-fixed path integral (\ref{170}) with (\ref{180}) and
(\ref{185})
for the triangular lattice which takes  into account
all defect degrees of freedom.
\end{appendix}

\end{document}